\documentclass[12pt]{iopart}

\usepackage{iopams}
\usepackage[dvips]{graphicx}

\begin{document}

\title{Dynamics of a stored Zeeman coherence grating in an external magnetic field}

\author{D. Moretti, D. Felinto, J. W. R. Tabosa}
\address{Departamento de F\'isica, Universidade Federal de
Pernambuco, Cidade Universit\'aria, 50670-901 Recife, PE, Brasil}

\author{A. Lezama}
\address{Universidad de la Rep\'ublica, C.P. 30 Montevideo
11000, Uruguay}

\date{\today}
\ead{tabosa@df.ufpe.br}

\begin{abstract}
We investigate the evolution of a Zeeman coherence grating induced
in a cold atomic cesium sample in the presence of an external
magnetic field. The gratings are created in a three-beam light
storage configuration using two quasi-collinear writing laser pulses
and reading with a counterpropagating pulse after a variable time
delay. The phase conjugated pulse arising from the atomic sample is
monitored. Collapses and revivals of the retrieved pulse are
observed for different polarizations of the laser beams and for
different directions of the applied magnetic field. While magnetic
field inhomogeneities are responsible for the decay of the coherent
atomic response, a five-fold increase in the coherence decay time,
with respect to no applied magnetic field, is obtained for an
appropriate choice of the direction of the applied magnetic field. A
simplified theoretical model illustrates the role of the magnetic
field mean and its inhomogeneity on the collective atomic response.

\end{abstract}

\maketitle

\section{Introduction}

The mapping of optical information into an atomic ensemble, a
phenomenon usually named light storage (LS), plays a fundamental
role for both classical and quantum information processing
\cite{Lukin03}. To date several observations of LS have been
reported both in thermal and in cold atomic systems
\cite{Phillips01,Liu01,Zibrov02,Wang05,Tabosa07, Moretti08}. Long
storage times as well as the possibility of controlling and
manipulating the stored information \cite{Pan09, Kuzmich09}, is of
essential importance for the realization of any practical quantum
protocol. In most of the experimental observations of LS, ground
state hyperfine levels of alkali atoms are employed to store the
optical information into Zeeman coherences. These coherences may be
conveniently manipulated by external magnetic fields \cite{Mair02,
Kumarakrishnan08, Peters09}. However, they are also very sensitive
to stray magnetic fields gradients, which can strongly reduce the
coherence time of the system \cite{Garreau01, Felinto05}. Collapses
and revivals due to Larmor precession of Zeeman coherence associated
with different hyperfine ground states of rubidium atoms have
already been observed, both using Poissonian classical light and
single photon quantum states \cite{Matsukevich06, Jenkins06}.
Recently, we have demonstrated the coherent Larmor precession of
orbital angular momentum of light stored into Zeeman coherences of
cesium atoms \cite{Moretti09}. It is also worth mentioning the
recently demonstrated long-lived gradient echo memory, which is a
different technique for LS and allows the control of the retrieved
pulse sequence \cite{Hetet08, Hosseine09}.

In this work we investigate the coherent evolution of a light
grating stored onto the ground state Zeeman coherence in the
presence of an external dc magnetic field. The presence of the
magnetic field give rise to a periodic evolution of the atomic state
(Larmor precession) that determines an oscillatory response of the
retrieved pulse efficiency as a function of the time separation
between the writing and reading of the coherence grating.
Nevertheless, the coherent response of the sample is damped as a
consequence of the dephasing of the individual atomic contributions
due to residual magnetic field inhomogeneities. Interestingly
enough, the damping time of the total atomic response strongly
depends on the direction of the applied magnetic field. We have
observed a five-fold increase in the coherence decay time, with
respect to the damping time at zero applied field, for a particular
direction of the applied magnetic field.

Precise theoretical modelling of our observations is prevented by
the lack of a priori knowledge of the residual magnetic field in the
sample. Nevertheless, the essential features of the experimental
results are qualitatively reproduced by a simplified theoretical
model, based on optical Bloch equations, describing the atomic state
evolution in the presence of a homogeneous magnetic field together
with an ad hoc inhomogeneous  field. This model accounts for the
oscillations (revivals) of the retrieved pulse amplitude as well as
for the variation of the storage decay time with magnetic field
direction. The key role played by the relative orientations of the
mean magnetic field and its inhomogeneity is also illustrated
through the classical calculation of the precession of an ensemble
of magnetic dipoles in the presence of an inhomogeneous field.

\section{Experimental setup and results}
\label{Experimental setup and results}

The experimental apparatus is similar to the one previously
described in \cite{Moretti08} and employs cold cesium atoms obtained
from a magneto-optical trap (MOT). The atoms are initially prepared
into the $6S_{1/2}(F=3)$ hyperfine ground state by optical pumping
induced by the MOT trapping beams, with the repumping beam switched
off for about 1 ms. The MOT quadrupole magnetic field is switched
off during the optical pumping process and the subsequent light
storage sequence. Three pairs of orthogonal Helmholtz coils are used
to compensate for stray magnetic fields. To write and read the
coherence grating we use light from an external cavity diode laser,
locked to the cesium D2 transition $6S_{1/2}(F=3)\leftrightarrow
6P_{3/2}(F^{\prime }=2)$. Immediately after the optical pumping
interval, two laser beams are incident on the atomic cloud making a
small angle of $\theta = 60$ mrad as indicated in Fig. 1-(a). These
two beams, named from now on as the grating writing beams, $W$ and
$W^{\prime}$, have opposite circular polarizations and induce a
Zeeman coherence grating between pairs of Zeeman sublevels, as
indicated in the Fig.1 (a). The writing beams have their intensity
controlled by a pair of acousto-optic modulators (AOM) and are
applied during a time long enough ($\approx 40 ~\mu$s) so to allow
the system to reach a steady state regime. During the
coherence-grating writing interval, an external dc magnetic field is
applied, either along the propagation direction of the $W$ beam
(chosen as quantization axis in Fig. 1) or orthogonal to this axis,
by two additional independent pairs of Helmholtz coils shown in
Fig.1. After the switching off of the writing beams, the induced
Zeeman coherence grating evolves in the presence of the external
magnetic field undergoing Larmor precession. The instantaneous
Zeeman coherence grating can be used to retrieve the optical field
(the diffracted beam $D$) by switching on a reading beam, $R$,
counterpropagating to the writing beam $W$ and having a circular
polarization opposite to that of this beam, as shown in Fig. 1-(b).
For this beam geometry, the retrieved light pulse is phase
conjugated to the writing beam $W^{\prime}$ and couterpropagating
with respect to this beam. The reading beam also passes through an
independent pair of AOMs that allows for the variation of the time
interval between writing and reading processes as indicated in the
time sequence shown in Fig. 1-(c). We have experimentally verified
that the retrieved beam has a polarization which is orthogonal to
the polarization of the reading beam as indicated in Fig. 1-(b).

The  retrieved signal intensity for different storage times is shown
in Fig.2-(a) corresponding to the polarization configuration
specified in Fig. 1-(a),(b) for the case where no external dc
magnetic field is applied. For these data, the intensities of the
writing beams $W$ and $W^{\prime}$ are equal to 17.6 mW/cm$^{2}$ and
1.1 mW/cm$^2$ respectively, while the intensity of the reading beam
is equal to 7.2 mW/cm$^2$. As can be observed, the retrieved signal
peak intensity has an approximately exponential decrease with a free
decay time of the order of 4 $\mu$s \cite{Moretti08}. This measured
coherence decay time is mainly determined by the non compensated
residual magnetic field gradients \cite{Felinto05}. The MOT
temperature was estimated to be in the range of mK and the measured
density of atoms in the F=3 ground state was of the order of $n
\approx 10^{10}$ cm$^{-3}$. From these values, we estimated the
collision rate to be of order of $\Gamma_{coll} \approx  0.2$ KHz
\cite{Hopkins00}, therefore having a negligible effect in the
decoherence process. Also, at the estimated temperature, the time
needed for an atom to move one grating period is much larger than
the measured coherence time, therefore ruling out the effect of
atomic motion.

The dynamics of the retrieved signal changes completely when a dc
magnetic field of magnitude $B_{x}$=0.7 G is applied perpendicularly
to the propagation axis, i.e.,  along the $x$ direction. The
observed result is presented in Fig. 2-(b), which clearly shows a
periodic oscillation of the retrieval efficiency with a period of
approximately $T_{L}$=4.4 $\mu$s. On the other hand, as shown in
Fig. 2-(c), if the dc magnetic field is applied along  the $z$ axis,
we also observe a periodic modulation of the retrieved signal, but
now with smaller visibility and a period given approximately by
$T_{L}$/2=2.2 $\mu$s. These values are consistent with Larmor
frequency predicted for the hyperfine level $(F=3)$ , which is given
by $\Omega_{L}=g_{F}\mu_{B}B$, with $g_{F}=-1/4$ the Land\'e factor,
$\mu_{B}$ the Bohr magneton and $B$ the magnitude of the magnetic
field \cite{Steck}. From the applied magnetic field in our setup we
estimate a Larmor period of $T_{L}=2\pi/\Omega_{L}$=4.1 $\mu$s,
which agrees reasonably well with the observed value.

\vspace{0.2cm}
\begin{figure}[ht]
\begin{center}
\hspace{-0.0cm}
\includegraphics[angle=0,scale=0.5]{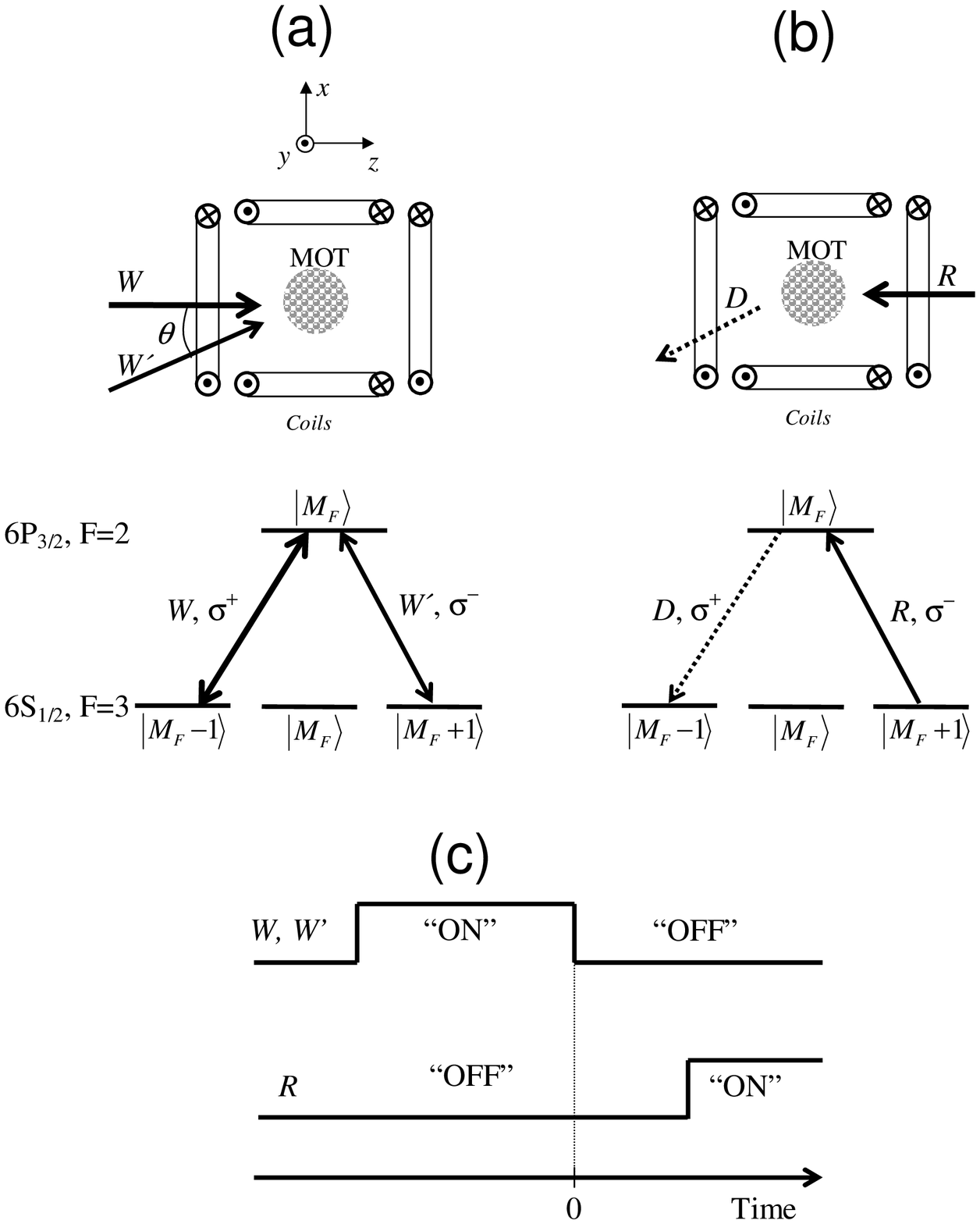}
\end{center}

\vspace{-2.0cm}

\caption{(a) Incident writing beams ($W$ and $W^{\prime}$)
configuration and partial Zeeman level scheme of the hyperfine
transition $(F=3)\leftrightarrow (F^{\prime }=2)$, showing the
coupling of the  writing beams. (b) Spatial configuration and
coupling of the incident reading ($R$) and the retrieved diffracted
($D$) beams . The beams $W$ and $W^{\prime}$ make a small angle
$\theta$ and are circularly polarized with opposite handedness,
while the beam $R$ is counterpropagating to the beam $W$ and have a
circular polarization opposite to this beam. The diffracted beam is
detected in a direction opposite to the beam $W^{\prime}$. (c)
Switching time sequence for the writing and reading fields.}
\end{figure}

\begin{figure}[ht]
\begin{center}
\hspace{-0.0cm}
\includegraphics[angle=0,scale=0.5]{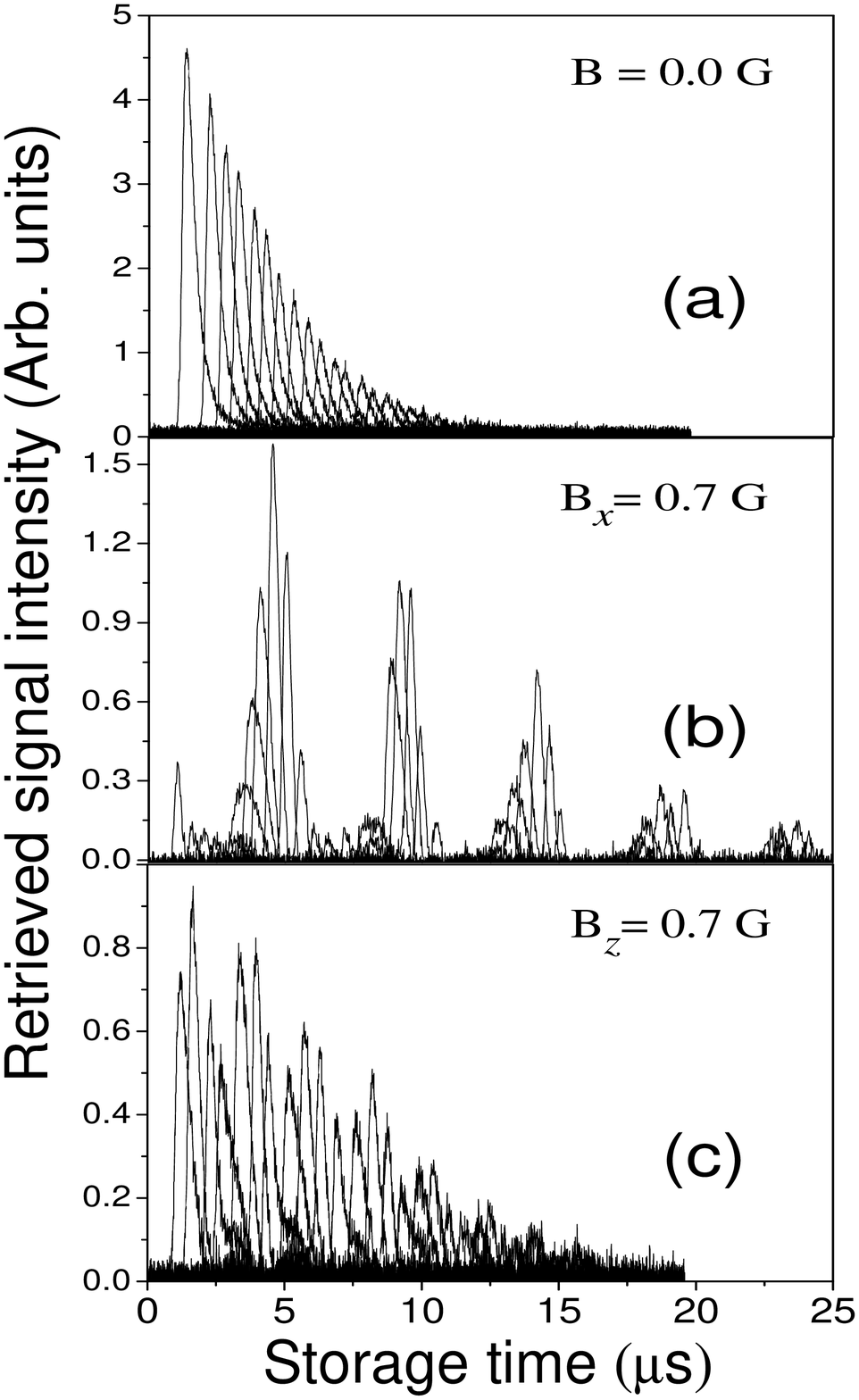}
\end{center}

\hspace{-0.5cm} \caption{Retrieved signal for different storage
times when: (a) no external dc magnetic field is applied; (b) a
$B_{x}$=0.7 G dc magnetic field is applied along the $x$ direction;
(c) a $B_{z}$=0.7 G dc magnetic field is applied along the $z$
direction. Each graphic represents a superposition of the retrieved
pulses associated with different reading time.}
\end{figure}

A simple picture can be given for a qualitative understanding of the
observed behavior by considering the evolution of the coherent
superposition of Zeeman sublevels, prepared by the writing beams, in
the presence of the applied magnetic field. Considering the larger
intensity of beam $W$ with respect to $W^{\prime}$, one should
expect orientation of the atomic system with the largest population
in the Zeeman sublevel with the highest $M_{F}$ value. Thus, in the
former case where the magnetic field is perpendicular to the
quantization direction, the magnetic hamiltonian will couple the
Zeeman states differing by $\Delta M_{F}= \pm 1$ and the probability
amplitude associated to each component of the initial superposition
state will oscillate at the Larmor frequency, therefore repeating
itself after the Larmor period. It is worth noticing that after half
the Larmor period the evolution changes the probability amplitude
associated with a sublevel $M_{F}$ into that of $-M_{F}$ and
vice-versa, therefore also repeating the value of the Zeeman
coherence. However, at this instant the reading beam will be coupled
to a different transition, which can lead to a decrease of the
diffracted signal. Indeed, in the limit of strong saturation of the
reading beam, the signal is reduced by the ratio between the two
Clebsch-Gordan coefficients of the transitions coupled by the
reading in the two different situation \cite{Moretti08}.
Differently, when the dc magnetic field is applied along the
quantization direction, each component of the initial superposition
will simply acquire a phase factor proportional to
$exp(iM_{F}\hbar\Omega_{L})$ and the Zeeman coherences associated
with sublevels differing by $\Delta M_{F}= 2$ will therefore
oscillate with a period $T_{L}/2$. This simple consideration
accounts qualitatively well for the series of collapses and revivals
observed experimentally.

In addition, we have performed similar measurements for the case
where the writing beams have orthogonal linear polarization with the
reading beam being linearly polarized parallel to the polarization
of the writing beam $W^{\prime}$. In this case, alignment is
produced in the atomic system leading to a symmetric distribution of
the Zeeman population. In such case, for an applied transverse
magnetic field, after half the Larmor period all the Zeeman
coherences and populations repeat themselves and therefore the
retrieved signal oscillates with half the Larmor period. This
behavior was experimentally observed. Finally, we also have measured
the evolution of the Zeeman coherence grating for different values
of the external magnetic field.  In Fig. 3 we plot the inverse of
the Larmor period as a function of the transverse magnetic field
amplitude. As expected, a linear dependence is observed.

The most remarkable feature of the measurements presented in Fig. 2
is the observed increase of the coherence decay time when the dc
magnetic field is applied. As mentioned before, the measured
coherence decay time is mainly determined by stray magnetic field
gradients that are not precisely known in our apparatus. The
experimental data shown in Fig.2(b) and Fig.2(c) clearly indicate
that a larger increase in the coherence time is obtained when a dc
magnetic field is applied along a specific direction ($x$ axis),
which is orthogonal to the propagation direction of $W$ and $R$
beams, although a smaller increase is also observed when the
magnetic field is applied along the propagation direction of these
beams ($z$ axis).

\begin{figure}[ht]
\begin{center}
\hspace{-0.0cm}
\includegraphics[angle=-90,scale=0.4]{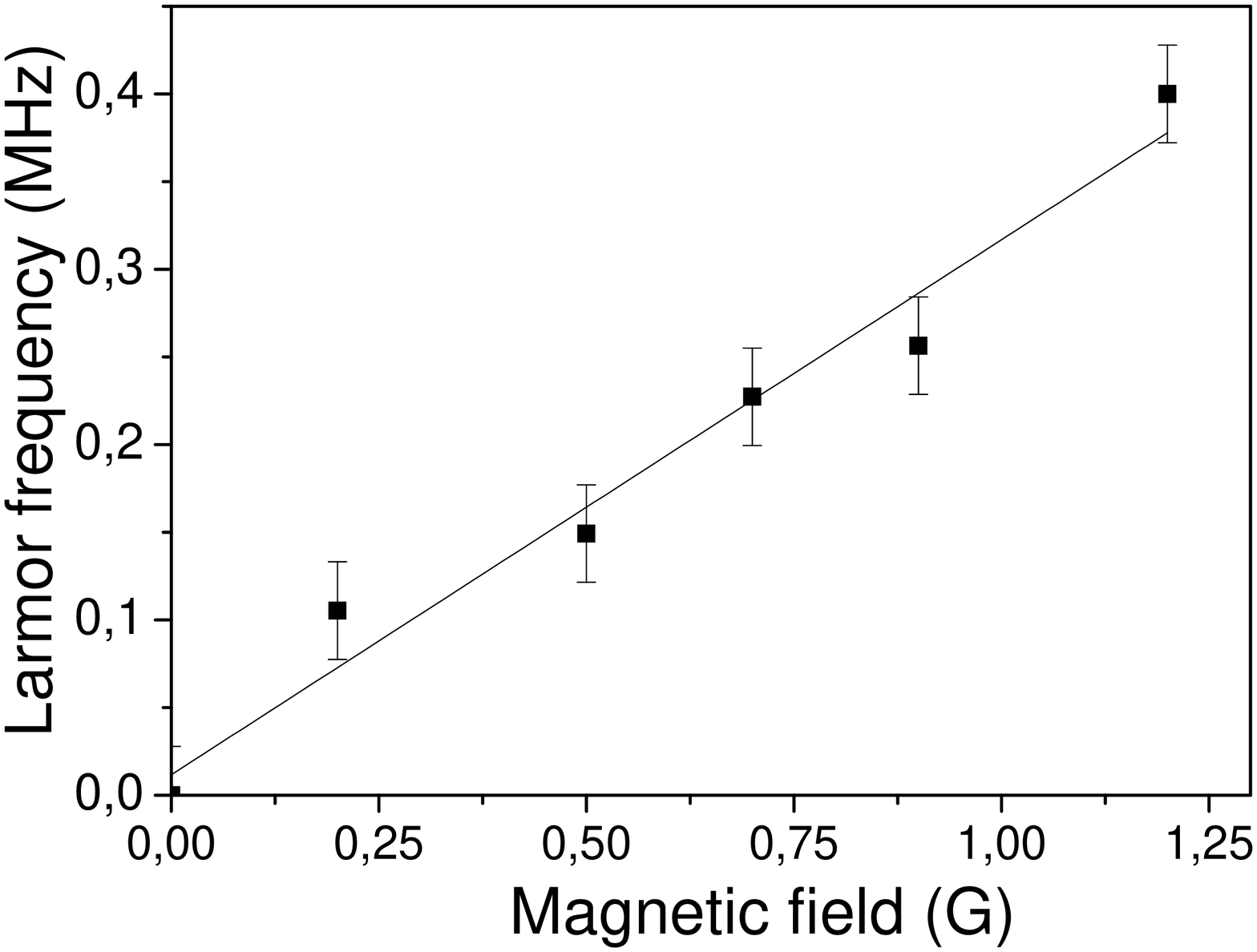}
\end{center}

\caption{Measured Larmor frequency as a function of the applied
transverse magnetic field. The solid line is a linear fitting.}

\end{figure}

\section{Theoretical modelling}
\label{Theoretical modelling}

In this section we provide a simplified model of the atomic
evolution in the presence of the writing and reading light pulses,
which gives rise to the emission of the retrieved pulse. Although
the model does not intend to precisely fit the experimental
observations, the main features of the observed signals are
qualitatively reproduced. The model gives a clear indication of the
relevant physical parameters determining the atomic evolution.

The model is based on the numerical solution of the Bloch equations
for an ensemble of atoms evolving in a dc magnetic field. The
calculation is a direct extension of the one presented in
\cite{Akulshin06}. The reader is referred to this publication for
details. We briefly remind here the main lines of the calculation.
Two atomic hyperfine levels $g$ and $e$ are considered with well
defined total angular momentum $F_{g}$ and $F_{e}$ respectively,
connected by an optical transition. All Zeeman sublevels are taken
into account. The excited state $e$ has a total decay rate $\Gamma$
that determines the relaxation of the excited state population and
of the optical coherence. Unlike in \cite{Akulshin06}, however, we
do not include in the model any relaxation mechanism for the ground
state coherence. The light field acting upon the atoms is resonant
with the atomic transition. The Bloch equations are numerically
integrated for successive time intervals, corresponding to the light
storage sequence with square pulses. Each interval corresponds to
the duration of an applied light pulse. During each interval, the
optical field has a well defined amplitude, polarization and phase.
In a given time interval, the optical field may correspond to a
unique ($R$) light pulse or be the total field resulting from two
light beams ($W$ + $W^{\prime}$).

Our simplified model does not include light propagation effects
across the atomic sample. The atoms are subject to a single optical
field, which can however have one or two polarization components. In
consequence, no spatial grating is represented in the model.
Numerically integrating the Bloch equations, we calculate the atomic
polarization during the reading time-interval. We consider the beams
$W$ and $W^{\prime}$ with orthogonal linear polarizations. Beams
$W^{\prime}$ and $R$ have the same polarization. Notice that the
choice of linear polarizations is different from the experimental
situation previously described. The reasons for such choice are
explained next.

We are interested in the  transient atomic response corresponding to
the emission of the ``phase conjugate" retrieved pulse. In the
experiment such pulse is characterized by its propagation direction
opposite to that of field $W^{\prime}$. Since our calculation does
not include propagation, phase matching effects determining the
emission of light in well defined directions are not accounted for
and cannot be used to characterize the ``phase conjugate" retrieval.
In consequence,  only polarization  can be used in the calculation
for the identification of the ``phase conjugate" retrieved pulse.
The corresponding light intensity is proportional to the square
modulus of the optical coherence of the atomic sample with
polarization opposite to that of fields $R$ and $W^{\prime}$.

The fact that in the numerical simulation, only the light
polarization can be used to separate the ``phase conjugate"
retrieved light pulse, imposes some constraints in the choice of the
light polarizations. In particular, we have to be careful to choose
the common polarizations of fields $R$ and $W^{\prime}$ such that it
will not acquire a component along the ``detected" polarization (the
same as $W$) as a consequence of Faraday rotation under the dc
magnetic field. Such effect, that would contaminate the retrieved
pulse with a background independent of the previously stored light
pulses, is not relevant in the experiment where only the phase
conjugate response is phase matched in the direction opposite to
that of field $W^{\prime}$.

We have considered in the numerical simulation a linear polarization
along axis $x$ for field $W$, and linear polarization along $y$ for
fields $W^{\prime}$ and $R$. The ``phase conjugate" retrieved pulse
is due to the atomic polarization during the reading interval
oriented along $x$. Two orientations of the mean  magnetic field
were considered: $x$ and $y$.

In cold atoms, where atomic motion and collisions can be neglected,
the decay of the coherent response of the atomic sample comes from
dephasing of the contributions of atoms in different parts of the
sample. Such dephasing is due to the presence of an inhomogeneous
magnetic field along the sample that causes different evolutions of
the atomic ground state \cite{Garreau01, Felinto05}. In order to
simulate the inhomogeneous response of the atomic sample, we have
calculated the evolution of homogeneous atomic sub-samples  subject
to randomly distributed magnetic fields. The total response is given
by the sum of the optical coherence in each sub-sample. We have
considered in our calculation different orientations of the mean
value $\textbf{B}$ of the  vector magnetic field. However, in order
to simplify the calculation and illustrate the importance of the
inhomogeneity orientation, we have only considered magnetic field
departures from the mean value  with a constant orientation (along
$x$). A Gaussian distribution of the magnetic field variation was
assumed with variance $\Delta B$.

\begin{figure}[ht]
\begin{center}
\hspace{-0.0cm}
\includegraphics[angle=-90, scale=0.4]{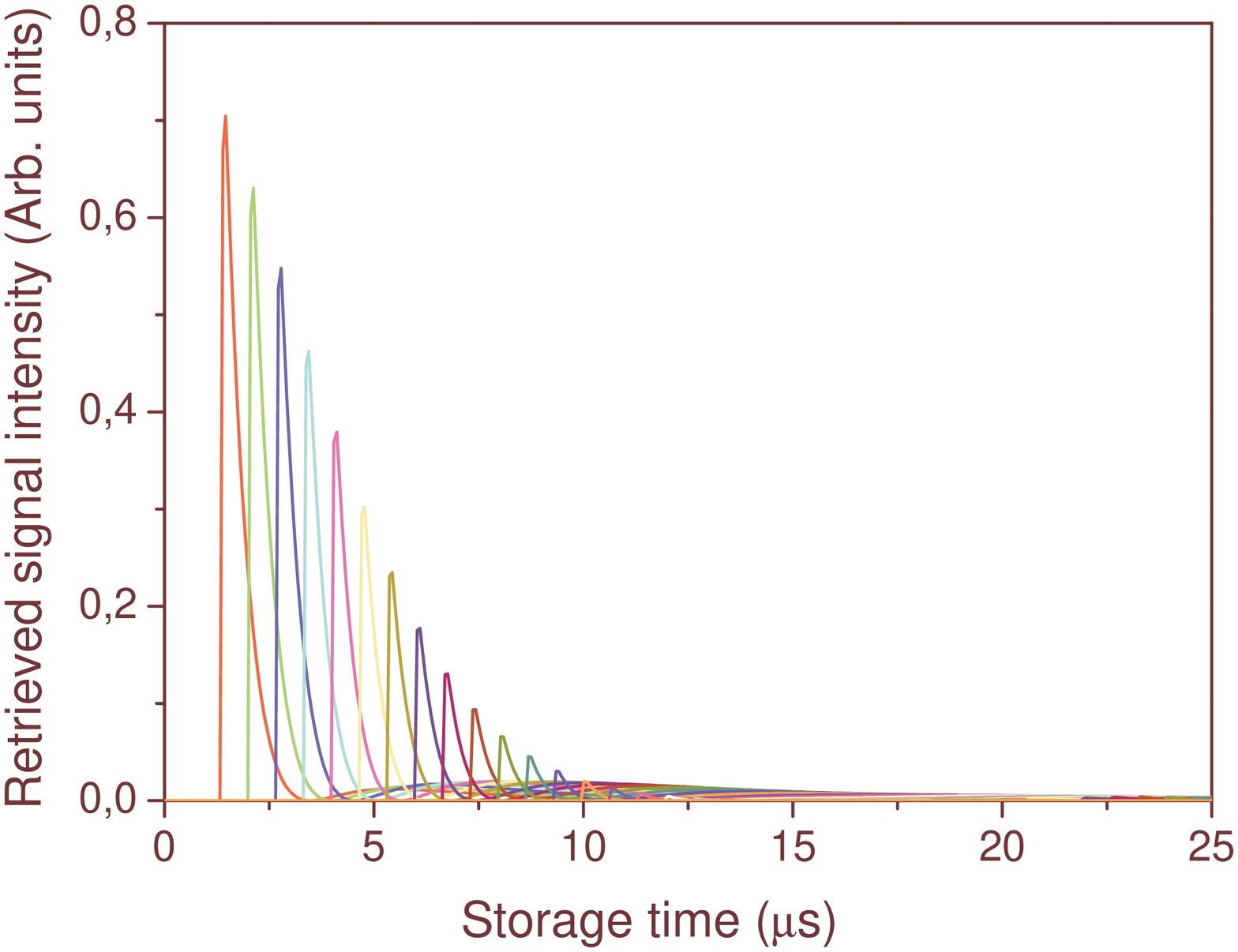}
\end{center}

\caption{\label{campo0} (Color online) Calculated retrieved light
pulses for different storage times for an ensemble of atoms in an
inhomogeneous magnetic field oriented along axis $x$ with zero mean.
The  time origin corresponds to the simultaneous turn off of the
writing fields. Each trace initiates at a time corresponding to the
turn on of the reading field. }
\end{figure}

\begin{figure}[ht]
\begin{center}
\hspace{-0.0cm}
\includegraphics[angle=-90, scale=0.4]{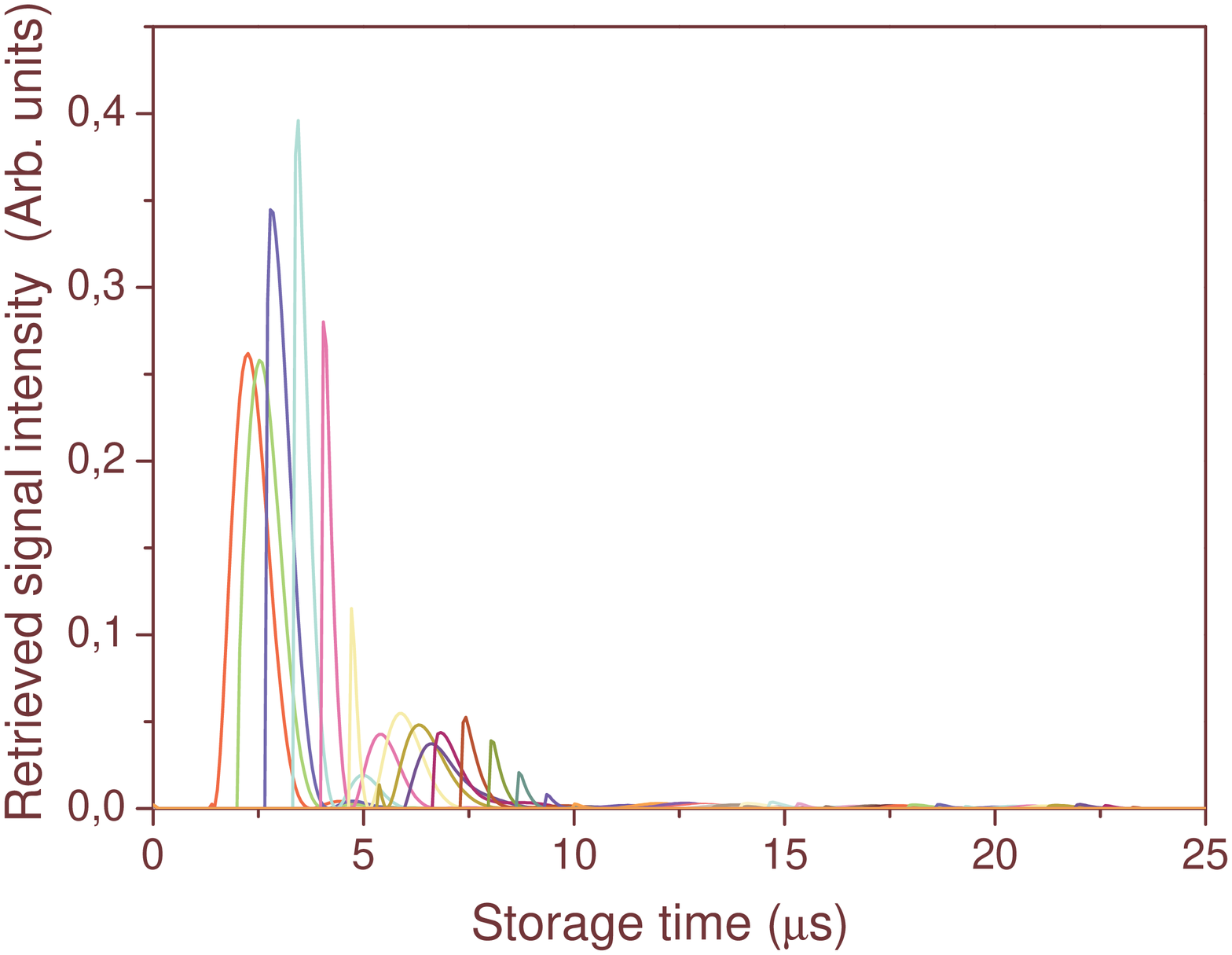}
\end{center}

\caption{\label{campox} (Color online) Same as Fig. \ref{campo0} for
an ensemble of atoms in the presence of a mean magnetic fields
oriented along $x$ in addition to a random distribution of magnetic
fields oriented along axis $x$ with zero mean. }
\end{figure}

\begin{figure}[ht]
\begin{center}
\hspace{-0.0cm}
\includegraphics[angle=-90,scale=0.4]{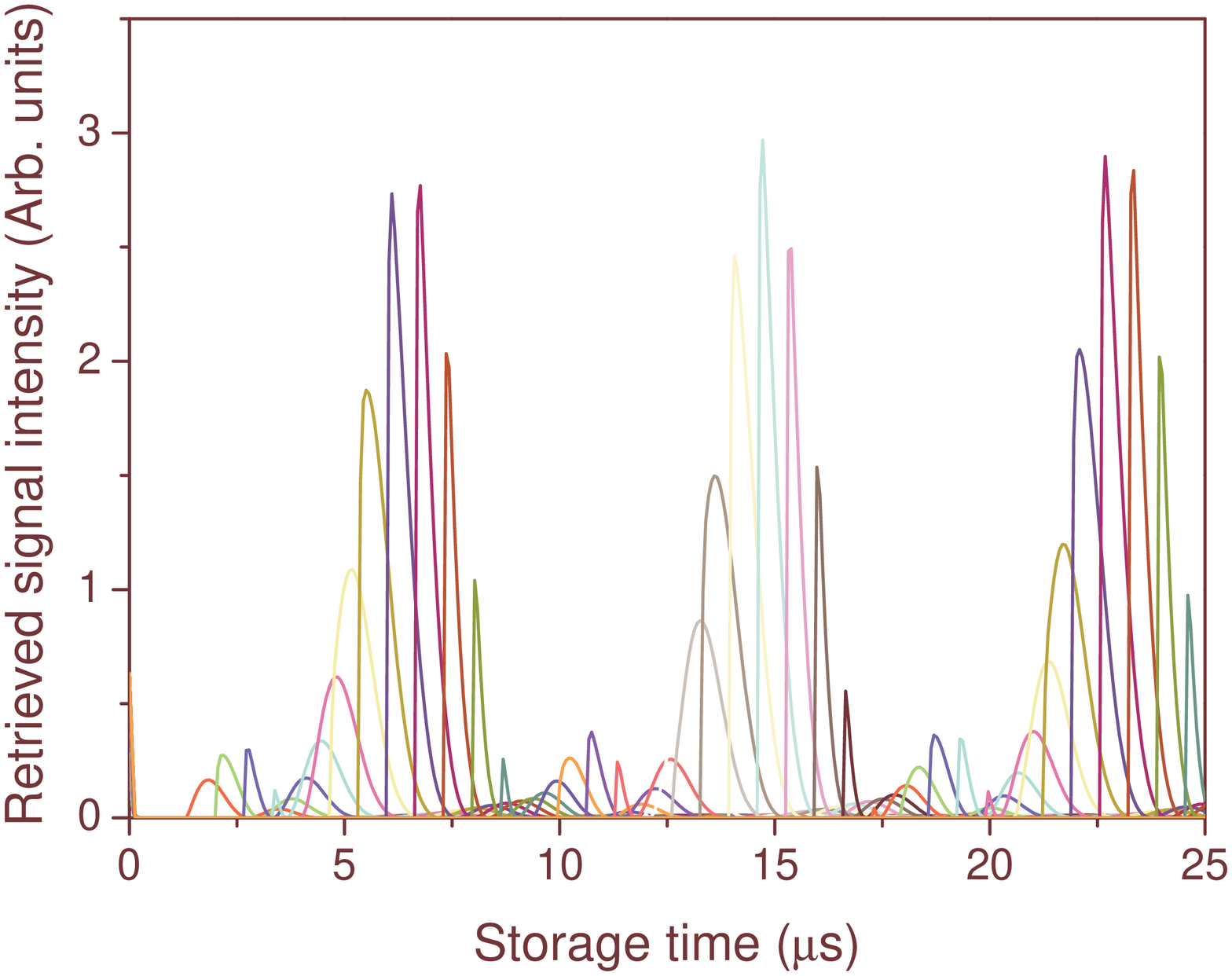}
\end{center}

\caption{\label{campoy} (Color online) Same as Fig. \ref{campo0} for
an ensemble of atoms in the presence of a mean magnetic fields
oriented along $y$ in addition to a random distribution of magnetic
 fields oriented along axis $x$ with zero mean.
}
\end{figure}

Figures \ref{campo0}-\ref{campoy} show the result of the numerical
simulations of the retrieved ``phase conjugate" light pulse as a
function time. Although the simulation reproduces the complete light
storage sequence, only the storage and retrieval intervals are
represented. The origin of the time axis corresponds to the
simultaneous turn off of the two writing fields $W$ and
$W^{\prime}$. In each figure, several retrieved pulses are presented
corresponding to different durations of the dark (storage) interval.
For simplicity, the calculations were carried on a model transition
$F_{g}=1 \rightarrow F_{e}=0$. Other choices of the angular momenta
give qualitatively similar results. The amplitudes of the applied
fields are determined by the corresponding reduced Rabi frequency
for the transitions. The simulations presented in Figs.
\ref{campo0}-\ref{campoy} correspond to reduced Rabi frequencies of
$0.5\Gamma$,  $0.25\Gamma$ and $0.125\Gamma$ for fields $W$,
$W^{\prime}$ and $R$, respectively. The duration of the writing
period ($W$ and $W^{\prime}$ both present) was taken as $10.6\mu s$

Figure \ref{campo0} show the retrieved pulses obtained for zero mean
magnetic field in the presence of a Gaussian distributed
inhomogeneous magnetic field oriented along $x$ with variance
 $g_{F}\mu_{B}\Delta B=5\times 10^{-3}\Gamma$. The magnetic field inhomogeneity is responsible for
 the observed decay of the retrieved pulse amplitude with a characteristic decay time of the
 order of a few microseconds.

When, in addition to the inhomogeneous magnetic field distribution,
a dc magnetic field with $g_{F}\mu_{B} B=2\times 10^{-2}\Gamma$ is
applied along axis $x$ an oscillation is present at twice the
corresponding Larmor frequency (Fig. \ref{campox}). The envelope
decay time of the retrieved pulses amplitude oscillation is similar
to that in Fig. \ref{campo0}. Notice that the shape of the
individual pulses varies depending on whether the retrieval process
occurs during the rising or the falling of the magnetically induced
oscillation. Broad pulses are observed on the rising side of the
oscillation and narrow ones on the falling side. A similar effect is
observed in the experiment (See Fig. 2). In the presence of the
magnetic field, individual retrieved pulses also present
oscillations that are rapidly damped with a decay time dependent on
the $R$ pulse intensity.

A dramatically different behavior is seen in Fig. \ref{campoy} where
the mean magnetic field is oriented along axis $y$ while the field
inhomogeneity is given by field variations along $x$. In this case,
the oscillation of the peak of the retrieved pulse is clearly
visible while the envelope decay is almost unnoticeable on the time
scale of the figure. Notice the presence of two frequency components
in the oscillation: a larger amplitude oscillation at the Larmor
frequency and a smaller oscillation at twice that frequency.

\begin{figure}[ht]
\begin{center}
\hspace{-0.0cm}
\includegraphics[angle=-90, scale=0.6]{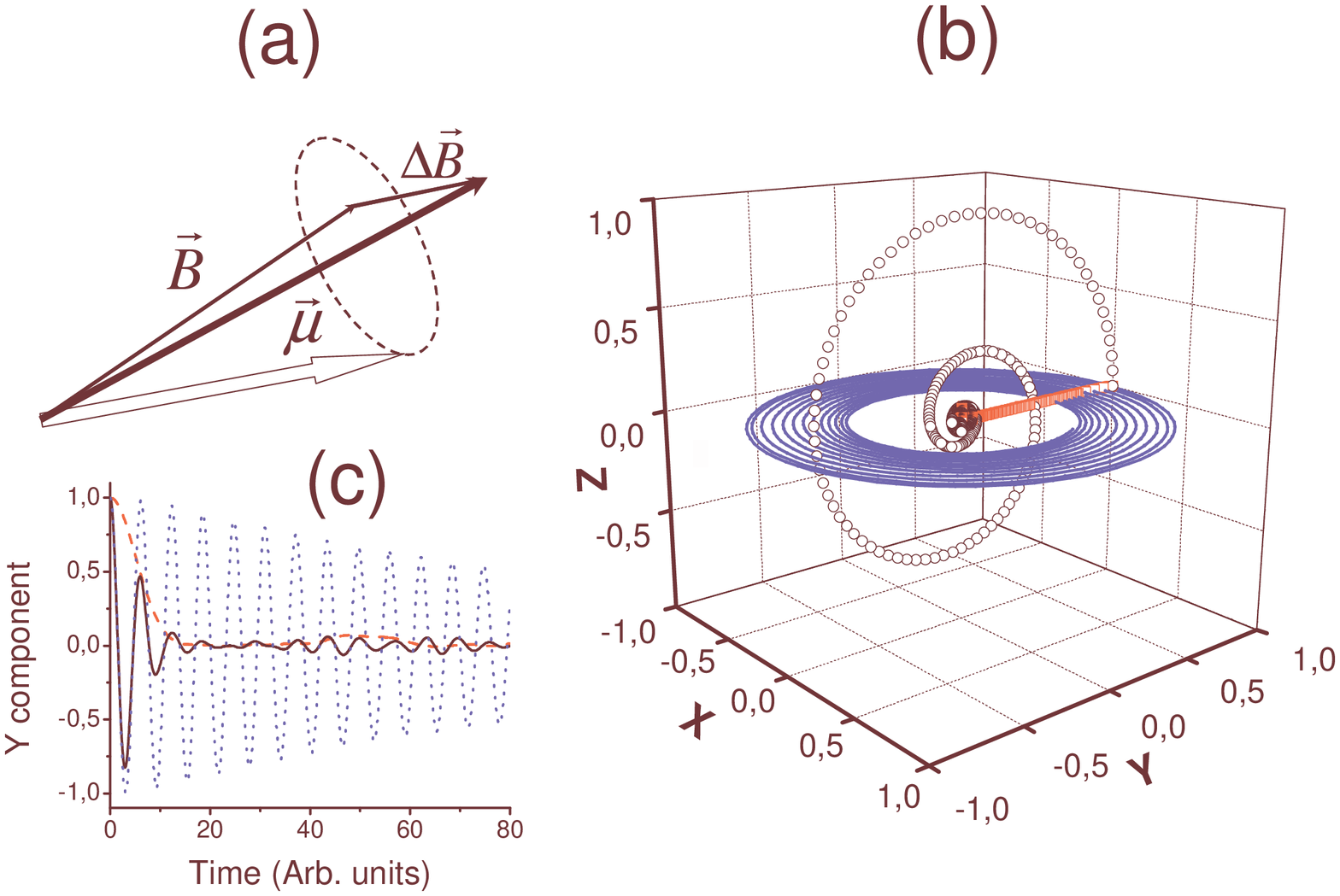}
\end{center}

\hspace{-1.0 cm}

\caption{\label{classico} Color online. a) Precession of a magnetic
dipole $\vec{\mu}$ around a magnetic field. b) Trajectory of the
total magnetic moment for an ensemble of magnetic dipoles for
different orientations of the mean magnetic field: no mean field
(hollow red squares); mean field along $x$ (hollow black circles),
mean field along $z$ (solid blue line). The magnetic field
variations with respect to the mean value are oriented along $x$. c)
$y$ axis projections of the total magnetic moment: no mean field
(dashed red); mean field along $x$ (solid black), mean field along
$y$ (dotted blue). }
\end{figure}

Figure \ref{campoy} dramatically illustrates the importance of the
relative orientation of the mean magnetic field and the field
inhomogeneity for the effective decay rate of the retrieved pulse
amplitude. When the mean field orientation is perpendicular to the
inhomogeneous field, the dephasing of the contributions to the
signal from atoms at different locations occurs at a much slower
rate. This effect has a purely classical origin as can be
illustrated by considering the (classical) motion of an ensemble of
magnetic dipoles precessing in a homogeneous magnetic field. We
present in Fig. \ref{classico} the evolution of the total magnetic
dipole, initially oriented along $y$ evolving in a magnetic field
with inhomogeneity in the $x$ component. If the mean magnetic field
is zero or oriented along $x$ a relatively short decay of the total
dipole is observed. For a mean magnetic field oriented along $z$ the
evolution of the total dipole corresponds to a slowly damped
precession. This difference in dephasing time can be simply
understood by noting that while an inhomogeneous field oriented
along the mean field will produce a first order correction to the
total field, a magnetic field variation in a direction perpendicular
to the mean field only causes a second order total field
modification. The same argument can be used for the atomic evolution
considered previously.

\section{Discussion and Summary}
\label{Discussion and Summary}

In spite of the simplicity of the theoretical model presented above
and the use of a different choice of optical polarizations, several
features of the experimental observations are qualitatively
reproduced. An almost exponential decay of the retrieved pulse
amplitude is observed in the absence of a mean magnetic field
illustrating the inhomogeneous nature of the coherence decay. In
this case all retrieved pulses have the same shape with a
characteristic decay time dependent on the reading pulse intensity.
When a dc magnetic field is present, the retrieved pulse amplitude
oscillates at the Larmor frequency and/or its second harmonic. This
is consistent with the expected evolution of the sample orientation
or alignment, respectively. In the general case, both frequencies
are present in the evolution of the atomic coherence. The individual
retrieved pulses have different shapes depending on the turn-on time
of the reading pulse with respect to the magnetic oscillation cycle.
Most importantly, the similarities between experimental results and
the calculated evolution illustrates the decisive role of the
magnetic field inhomogeneity on the coherence signal decay. When the
dc field is applied perpendicular to orientation of the local
variations of the magnetic field a significant enhancement in the
coherence decay can be observed. Such feature may be used for the
characterization of unknown field inhomogeneities.

The above result can be seen as a simple means of coherence control
in a sample where the decoherence  is a consequence of spatial
inhomogeneity of the atomic sample. Coherence control has attracted
considerable attention in recent times due to its crucial role in
the preservation of quantum information. In the case where
decoherence results from the coupling of the observed system to a
large environment, and is thus homogeneous and irreversible in
nature, coherence control poses a really challenging problem.
Significant progress has been recently achieved in this field both
with passive and active control protocols relying on the shaping of
the interaction of the small system with its environment
\cite{Gordon07}. In the present case, however, the decoherence time
for the state of individual atoms is long and the shorter observed
decay of the coherent emission is the result of the dephasing of the
individual atomic contributions evolving in slightly different
magnetic environments. We have shown here that the resulting
effective dephasing time can be manipulated by the addition of a
constant magnetic field properly oriented. However, one has to keep
in mind that inhomogeneous dephasing is not intrinsically
irreversible. The initial information stored in the sample is, in
principle, available in the atomic sample where individual atoms
undergo unitary evolution. The stored information can be recovered,
by inducing the rephasing of the atomic contributions through photon
echoes \cite{Abella66}. Moreover, the inhomogeneous nature of the
atomic dephasing allows the simultaneous recording in the sample of
different light pulse Fourier components. As first suggested by
Mossberg \cite{Mossberg82}, this allows the storage and recording of
light pulses with complex temporal profiles. While early experiments
used the pre-existing sample inhomogeneity, an interesting recent
improvement consists in the at-will manipulation of the sample
inhomogeneity by the application of a user controlled external
magnetic field gradient. In such case, storage as well as
manipulation (delay and reversal) of complex pulse temporal
information is possible \cite{Hetet08,Hosseine09}. Finally, it is
worth emphasizing the significant progress recently achieved in the
direction of suppressing the inhomogeneous dephasing altogether by
eliminating the influence of the magnetic field through the use of
magnetically insensitive two-photon transitions (clock transitions),
where storage time of the order of millisecond have been observed
\cite{Pan09,Kuzmich09}.

In summary, we have experimentally observed the Larmor precession of
a coherence grating stored into the Zeeman sublevels of the cesium
$6S_{1/2}(F=3)$ ground state in the presence of an applied magnetic
field. Collapses and revivals of the retrieved signal with distinct
Larmor oscillation frequency were observed for different
distributions of Zeeman populations and coherences associated with
different polarization configuration of the writing beams. An
appreciable increase in the coherence decay time was observed for a
specific direction of the applied magnetic field. We have also
presented a simple theoretical model based on the numerical solution
of the Bloch equation including the Zeeman degeneracy which
qualitatively agrees with the main experimental observations.
Moreover, we have shown that the increasing in the coherence time
has a even simpler and purely classical explanation. Finally, we
think that the possibility of controlling the coherence time in a
larger ensemble of atoms can be of considerable interest for several
applications and specially for quantum information.

We gratefully acknowledge Marcos Aurelio for his technical
assistance during the experiment. This work was supported by the
Brazilian Agencies CNPq/PRONEX, CNPq/Inst. Mil\^{e}nio and FINEP.

\end{document}